\documentclass[12pt]{article}

\usepackage{amsmath}
\usepackage{amsthm}
\usepackage{amsfonts}
\usepackage{graphicx}
\usepackage{lscape}
\usepackage{rotating}
\usepackage{longtable}
\usepackage{extarrows}
\usepackage{float}
\usepackage{hyperref}
\usepackage{geometry}
\usepackage{tikz-feynman}
\usepackage{tikz}
\usepackage{phiman}
\usepackage{authblk}
\usetikzlibrary{shapes.misc}
\usetikzlibrary{decorations.pathmorphing}
\tikzset{snake it/.style={decorate, decoration=snake}}
\tikzset{cross/.style={cross out, draw=black, minimum size=2*(#1-\pgflinewidth), inner sep=0pt, outer sep=0pt},
cross/.default={8pt}}

\title{Chaos in a quantum rotor model}

\author[1]{Gong Cheng}
\affil[1]{Condensed Matter Theory Center and Department of Physics, University of Maryland, College Park, MD 20742, USA}
\author[1,2,3]{Brian Swingle}
\affil[2]{Maryland Center for Fundamental Physics and Joint Center for Quantum Information and Computer Science, College Park, MD 20742, USA}
\affil[3]{Institute for Advanced Study, Princeton, NJ 08540 USA}

\date{}

\begin{document}
\maketitle

\begin{abstract}
We study scrambling in a model consisting of a number $N$ of $M$-component quantum rotors coupled by random infinite-range interactions. This model is known to have both a paramagnetic phase and a spin glass phase separated by second order phase transition. We calculate in perturbation theory the squared commutator of rotor fields at different sites in the paramagnetic phase, to leading non-trivial order at large $N$ and large $M$. This quantity diagnoses the onset of quantum chaos in this system, and we show that the squared commutator grows exponentially with time, with a Lyapunov exponent proportional to $\frac{1}{M}$. At high temperature, the Lyapunov exponent limits to a value set by the microscopic couplings, while at low temperature, the exponent exhibits a $T^4$ dependence on temperature $T$.

\end{abstract}

\tableofcontents

\section{Introduction}
\subsection{Motivation}

There have been a number of recent developments in the field of many-body quantum chaos driven in part by newly found relations to other areas of physics. Connections to quantum information scrambling and to the black hole information problem via holographic duality have been particularly fruitful~\cite{Shenker2014,Kitaev_talk, Hayden2007, Sekino2008, Hosur2016}. On the theoretical side, this has led to an intense effort to compute so-called out-of-time-order correlators (OTOCs) in many-body systems~\cite{larkin1969quasiclassical,Almheiri:2013hfa, Nahum2017a, VonKeyserlingk2017, Xu2018, Sachdev1993, Gu2017, Luitz2017, Bohrdt2017a, Heyl2018, Lin2018, Nahum2017, Rakovszky2017, Khemani2017, Khemani2018,Lashkari2012, Shenker2014a,Aleiner2016, Swingle:2016jdj, Grozdanov2018,Patel2017,Swingle2018Resilience,Dressel2018SCweak,Menezes2018Sonic,Scaffidi2017Semiclassical}. On the experimental side, following a number of proposals~\cite{swingle2016,Zhu2016,Yao2016a,Halpern2016,Halpern2017,Campisi2017,Yoshida2017}, at least six early experiments have already been carried out~\cite{Garttner2016, Wei2016, Li2017a, Meier2017,Landsman2018, Wei2018}. The study of a special fermionic system with infinite-range random interactions, the Sachdev-Ye-Kitaev (SYK) model \cite{Sachdev:2015efa,Kitaev_talk,Maldacena:2016hyu}, has revealed interesting features about scrambling, conformal symmetry and holography.

Here we study many-body quantum chaos in another infinite-range model consisting of a large $N$ number of interacting $M$-component quantum rotors~\cite{sachdev}. Our motivations for this study are as follows. We are interested to understand if the maximal chaos of the SYK model can be replicated in other physical systems, particularly in spin models of a type amenable to experimental realization. Since maximal chaos in SYK and two-dimensional gravity have to do with a certain pattern of conformal symmetry breaking, it is natural to study other infinite-range models with quantum critical points to attempt to extract a minimal set of ingredients for maximal chaos. This is particularly relevant given the recent experimental developments, since it would be desirable to physically instantiate and experimentally study models with maximal chaos, but the SYK model is quite elaborate from an experimental perspective~\cite{2017PTEP.2017h3I01D,2018PhRvL.121c6403C}. Finally, the model we study has both paramagnetic and spin glass phases, and it would be interesting to understand the interplay between chaos and glassiness. Although we do not address the spin glass phase, our work represents a necessary first step to doing so.

The system is analyzed in the limit of large $M$ and $N$, where it is known to be solvable. One interesting features is the presence of both a disordered paramagnetic phase and a spin glass phase separated by second order phase transition. We study the paramagnetic phase with a particular focus on the region of the phase diagram near the transition. The physical observables of interest are related to scrambling and diagnose how quantum information stored in a local part of a system spread to the whole system through interactions. Quantitatively, this process can be measured by the growth of certain local operator in the Heisenberg picture, which is characterized by the thermal average of squared commutator of two operators at different locations. Focusing on the rotor variables, we study the following quantity,

\begin{equation}\label{eq0}
\mathcal{C}_{i,j}(t)=-\text{Tr}\left(\rho [n_i(t),n_j(0)]^2\right),
\end{equation}
where $n_i$ is the rotor at site $i$ and $n_i(t) = e^{i H t} n_i e^{-i H t}$ is the corresponding Heisenberg operator for Hamiltonian $H$. We specify the index structure of the $M$-component rotors more carefully below.

This is essentially a four-point function of rotor variables and can be calculated using perturbation theory. For some systems, this quantity grows exponentially with time $t$ for a period of time between the local relaxation time and the scrambling time where the commutator begins to saturate to its late-time value. This growth diagnoses chaos, and it shows how an initial local perturbation causes an influence that grows exponentially with time~\cite{larkin1969quasiclassical,Shenker2014,Kitaev_talk}. Accordingly, the growth exponent is called a quantum Lyapunov exponent. This notion has recently been generalized to define a whole spectrum of quantum Lyapunov exponents~\cite{2018arXiv180901671G}. In quantum systems satisfying a set of conditions related to thermalization, the exponent has an upper bound given by $\frac{2\pi T}{\hbar}$~\cite{boundonchaos}.		
		
\subsection{Model and results}
The model consists of a large $N$ number of $M$-component rotor fields, labeled by $\vec{n}_i$, with non-local and random interaction. The classical part of the Hamiltonian is
\begin{equation}
H_c=\sum_{(ij)}J_{ij}\vec{n}_i\cdot \vec{n}_j=\sum_{(ij)}\sum_{\mu=1}^M J_{ij}n_i^\mu n_j^\mu,
\end{equation}
where the $J_{ij}$ couplings Gaussian random variables with mean zero and variance $\frac{J^2}{N}$. The interaction is non-local in the sense that the summation over $(ij)$ runs over all pairs of rotors. In the quantum version of this model, a kinetic term is added for each rotor which is proportional to the conjugate angular momentum squared. The full quantum Hamiltonian is
\begin{equation}
H=\frac{g}{2}\sum_{i=1}^N \vec{L_i}^2+\sum_{(ij)}J_{ij}\vec{n}_i\cdot \vec{n}_j.
\end{equation}

This model is exactly solvable in the large $N$ and large $M$ limit \cite{sachdev}, defined by the following conditions:
\begin{equation}
\begin{split}
&\langle J_{ij}^2 \rangle=\frac{J^2}{N}\\
&\vec{n}_i\cdot \vec{n}_i=\sum_{\mu=1}^M (n_i^{\mu})^2=M,   \ \ \text{ for each $i$}.
\end{split}
\end{equation}
Here the angular brackets denotes disorder average, and more generally they denote a combination of disorder and quantum average. As shown in the previous work, in this limit a factor of $N$ can be factorized from the Euclidean Lagrangian. Therefore the path integral is dominated by its leading saddle point contribution. We briefly review this background material in the following.

The Euclidean path integral is
\begin{equation}\label{delta}
\begin{split}
Z=\int\prod_i D n_i(\tau) \delta(n_i(\tau)^2&-1)e^{-\int_0^\beta d\tau [\frac{1}{2g} \sum_{i=1}^{N}(\partial_{\tau} n_i)^2+\sum_{(ij)} J_{ij}n_i\cdot n_j]}\\
\rightarrow \int\prod_{i,a} D n_i^a(\tau) D \lambda_i^a(\tau)& \{e^{-\frac{1}{2g}\int_0^\beta d\tau  \sum_i\sum_{a=1}^q (\partial_{\tau} n_i^a)^2+\frac{J^2}{2N}\int_{0}^{\beta}d\tau \int_{0}^{\beta}d\tau' \sum_{(ij)} \sum_{a,b}  n_i^a(\tau)\cdot n_j^a(\tau)n_i^b(\tau')\cdot n_j^b(\tau')}\\ &\times e^{-\frac{1}{2}\int_0^{\beta} d\tau \sum_{i,a}\lambda_i^a n_i^a\cdot n_i^a}\}\\
=\int\prod_{i,a} D n_i^a(\tau) D \lambda^a(\tau)& \prod_{cd\mu\nu}D Q_{\mu\nu}^{cd}(\tau,\tau') e^{-NI_E(n_i^a,\lambda^a,Q_{\mu_\nu}^{cd})}
\end{split}
\end{equation}
where the final normalized Euclidean action is
\begin{equation}
\begin{split}
I_E=&\frac{1}{2g}\int_0^\beta d\tau\frac{1}{N}\sum_{i=1}^N\sum_{a=1}^q [\partial_{\tau} n_i^a(\tau)]^2+\frac{J^2}{2}\int_{0}^{\beta}d\tau\int_{0}^{\beta}d\tau'\sum_{ab\mu\nu} [\frac{1}{2}Q_{\mu\nu}^{ab}(\tau,\tau')^2-\\
&\frac{1}{N}\sum_{i=1}^N n_i^{a\mu}(\tau)n_i^{b\nu}(\tau')Q_{\mu\nu}^{ab}(\tau,\tau') ]+\frac{1}{2}\int_0^{\beta} d\tau \frac{1}{N}\sum_{i=1}^N\sum_{a=1}^q\lambda^a(\tau) n_i^a(\tau)\cdot n_i^a(\tau).
\end{split}
\end{equation}
In the second equality, the ensemble average of $J_{ij}$ has been taken using the replica trick. The upper index of $n^a$ represents the replica index ranging from $1$ to $q$, which is ultimately taken to be $q\rightarrow 0$. The delta functions enforcing the normalization are represented using auxiliary fields $\lambda_i^a$. Their dynamics is generated by quantum corrections which are suppressed by $\frac{1}{M}$. In the large $M$ limit, they are just numbers serving as a chemical potential. A factor of $N$ was extracted in the third step, so that the auxiliary field $Q$ is coupled to the site average of rotor pairs.

One can show that the saddle point is determined by
\begin{equation}
Q^{ab}_{\mu\nu}(\tau,\tau')=\frac{1}{N}\sum_{i=1}^N\langle n_i^{a\mu}(\tau) n_i^{b\nu}(\tau')\rangle.
\end{equation}
Treating $Q$ as a non-dynamical field, the action of the rotor fields is that of a free theory. Therefore the two-point correlator of rotor variables can be obtained exactly. It is known that this model has a paramagnetic phase (replica symmetric) and a spin glass phase (replica symmetry breaking) separated by a second order phase transition. Our main focus in this article is the chaotic behavior within the paramagnetic phase, especially near the critical point.

To analyze scrambling in the model, we deviate slightly from the strict large $N$ and large $M$ limit to account for the fluctuations of the two auxiliary fields $Q$ and $\lambda$ which are suppressed by $\frac{1}{N}$ and $\frac{1}{M}$, respectively. The squared commutator is obtained by taking $N$ much larger than $M$ and summing over all the terms that are proportional to $\left(\frac{t}{M}\right)^n$ at long time in the ladder diagrams, while only keeping the leading $\frac{1}{N}$ contribution. In this limit we will show that the squared commutator, Eq.~\eqref{eq0}, is proportional to $\frac{1}{NM}$ and the Lyapunov exponent is suppressed by $\frac{1}{M}$:
\begin{equation}
\frac{1}{M^2}\sum_{\mu\nu}\langle [n_i^{\mu}(t),n_j^{\nu}(0)]^2\rangle_\beta \ \ \sim \  \frac{1}{MN}e^{\frac{1}{M}f(T)t}.
\end{equation}
The chaos exponent is
\begin{equation}
\lambda_c = \frac{f(T)}{M}.
\end{equation}
In the above formula, $f(T)$ is a increasing function of temperature. It is proportional to $T^4$ for low temperature near the critical point, and it saturates to a value proportional to $\sqrt{Jg}$ at large $T$. Although the calculation is performed for leading $\frac{1}{M}$ expansion. The $T^4$ behavior at small $T$ is still true for finite value of $M$.

\section{Two-point function}

The analysis begins with the two-point function of rotor fields. At leading order of large $N$ and $M$ limit, the correlator can be obtained either from the previously mentioned method or by solving the Schwinger-Dyson (SD) equation. In the paramagnetic phase, one can use the time translation symmetry to write the SD equation diagrammatically as


%

\begin{align*}
\Sigma(i\omega_n)&=\sum_j\PMl{.-.}{j}+\sum_{jk}\PMl{.-.-.-.}{j}{k}{j}+\sum_{jkl}\PMl{.-.-.-.-.-.}{j}{k}{l}{k}{j}+\cdots\\
&=\sum_j\PMl{.-.}{j}+ \sum_j \PMl{-S-}{j} + \sum_{j} \PMl{-S-S-}{j}{j} + \cdots
\end{align*}
\begin{equation}\label{SDE}
\begin{split}
&\Sigma(i\omega_n)=\frac{J^2}{N}\sum_{j=1}^{N}G_j(i\omega_n)=J^2G(i\omega_n) \\
&G^{-1}(i\omega_n)=\frac{\omega_n^2}{g}+\lambda-\Sigma(i\omega_n).
\end{split}
\end{equation}

The two-point function is diagonal, $\langle n_i^\mu(\tau)n_j^\nu(\tau')\rangle =G(\tau,\tau')\delta_{ij}\delta_{\mu\nu}$, due to the $O(M)$ rotational symmetry and local $Z_2$ symmetry of the model. The solution of these equations is

\begin{equation}\label{solution}
G(i\omega_n)=\frac{2}{J}\left(\frac{\lambda}{4J}+\tilde{\omega}_n^2-\sqrt{\left(\tilde{\omega}_n^2+\frac{\lambda}{4J}\right)^2-\frac{1}{4}}\right).
\end{equation}
Note that the solution depends on frequency only through the dimensionless combination $\tilde{\omega}_n=\frac{\omega_n}{2\sqrt{gJ}}$. The chemical potential $\lambda$ is determined by the normalization condition
\begin{equation}\label{normalization}
G(\tau=0)=T\sum_n G(i\omega_n)=1.
\end{equation}

Now define the spectral function
\begin{equation}
\begin{split}
A(\omega)=&\,\frac{1}{\pi}\Im G(\omega+i0^+) \\
=&\,\text{sgn}(\omega)\frac{2}{\pi J}\sqrt{\frac{1}{4}-(\tilde{\omega}^2-\lambda')^2}, \ \ \ \lambda'=\frac{\lambda}{4J}.
\end{split}
\end{equation}
It is non-zero only when $-\frac{1}{2}+\lambda'<\tilde{\omega}^2<\frac{1}{2}+\lambda'$. The paramagnetic phase corresponds to $\lambda'>\frac{1}{2}$. The phase transition from paramagnetic phase to spin glass phase occurs at temperature $T_c$ when $\lambda'$ reaches $\frac{1}{2}$ and the system becomes gapless. From condition Eq.~\eqref{normalization} and $\lambda'=\frac{1}{2}$,  $T_c$ can be solved as a function of $J$ and $g$. In particular, near $T_c=0$, one obtains
\begin{equation}
g=\frac{9\pi^2}{16}J-\frac{\pi^2}{2}\frac{T_c^2}{J}+\cdots
\end{equation}

Sitting at the critical point, the two point function is proportional to
\begin{equation}\label{conformalrotor}
\sqrt{\frac{g}{J}}\frac{1}{\sin^2(\frac{\pi\tau}{\beta})}+\cdots
\end{equation}
at strong coupling, $\beta\sqrt{gJ}\gg 1$, and $\tau>0$.

We define the retarded Green's function as $G_R(\omega)=G(i\omega_n\rightarrow\omega+i0^+)$. It can be expressed using the spectral function
\begin{equation}
G_R(\omega)=-\int d\nu \frac{A(\nu)}{\omega-\nu+i0^+}.
\end{equation}

At the critical point, this expression allows us to obtain the real time dynamics by Fourier transformation,
\begin{equation}
\begin{split}
\langle [n_i^\mu(t),n_i^\mu(0)]\rangle \theta(t):=&(-i)G_R(t)\\=&\frac{2}{J}\frac{J_2(\tilde{t})}{\tilde{t}},
\end{split}
\end{equation}
where $J_2(z)$ is the Bessel function of the first class and $\tilde{t}$ denotes the normalized time $2t\sqrt{gJ}$. At large $\tilde{t}$ it has the asymptotic behavior
\begin{equation}
(-i)G_R(\tilde{t})\sim \tilde{t}^{-\frac{3}{2}}\cos\left(\tilde{t}-\frac{\pi}{4}\right).
\end{equation}
This is polynomially decaying at large $t$, and we will see in the next section that the four-point function constructed solely from it has no exponential growing behavior.

Chaos is obtained only after deviating slightly from the large $M$ limit. Since the kinetic term for the $\lambda$ field generated by quantum corrections is of order $\frac{1}{M}$, we need to take into consideration the $\lambda$ fluctuations to this order. Its correlation function is denoted $\langle \lambda(\tau)\lambda(0)\rangle$ and is given by
\begin{equation}
G_{\lambda}(\omega_n)=-\frac{1}{M}\frac{1}{\Pi(i\omega_n)}
\end{equation}
in frequency space. Here $\Pi(i\omega_n)$ is the polarization function given by
\begin{equation}
\Pi(i\omega_n)=T\sum_n G(i\nu_n)G(i\omega_n-i\nu_n)
\end{equation}

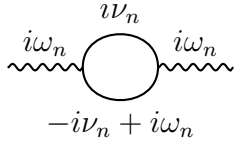
\begin{figure}
\centering
\feynmandiagram [layered layout, large, horizontal=b to c] {a -- [photon, edge label=\(i\omega_n\)]b -- [opacity=0.0]c-- [photon, edge label=\(i\omega_n\)] d, b -- [half left, looseness=1.5, edge label=\(i\nu_n\)] c -- [half left, looseness=1.5, edge label=\(-i\nu_n+i\omega_n\)] b, 
};
\caption{Diagram corresponding to the polarization function $\Pi(\omega_n)$. }
\label{bubble}
\end{figure}
Since there is a summation over $M$ rotor components in the loop, the propagator $G_\lambda(i\omega_n)$ is proportional to $\frac{1}{M}$. This justifies treating $\lambda$ as a non-dynamical field at large $M$ limit.

The polarization function is calculated in Appendix~\ref{lambdafield}. Generally, there is no analytical expression for $\Pi$, but at the critical point and in the strong coupling limit it has a simple form,
\begin{equation}\label{lambdareal}
 G_\lambda(\tau)=\frac{1}{M}\frac{c J^{2}}{\sin^4 \frac{\pi\tau}{\beta}}.
\end{equation}
It is also useful to define the spectral function of $\lambda$ as
\begin{equation}
\frac{1}{M}A_{\lambda}(\omega)=\frac{1}{\pi}\Im G_{\lambda}(\omega+i0^+)=-\frac{1}{\pi M}\Im\frac{1}{\Pi(\omega+i0^+)}.
\end{equation}

There is another two-point function, known as the Wightman function, that will be useful later. It can be obtained as via analytic continuation of the imaginary time correlation function,
\begin{equation}
G_W(t):=G\left(-\frac{\beta}{2}+it \right)
\end{equation}
One can derive a spectral function representation for $G_W(\omega)$ which reads
\begin{equation}
\begin{split}
G_W(t)&=T\sum_n \int dx \frac{-A(x)}{i\omega_n-x}e^{-i\omega_n(-\beta/2+it)}\\
&=\frac{1}{2\pi i}\int dx\oint dz\frac{A(x)}{z-x}n_B(z)e^{-z(-\beta/2+it)}\\
&=\frac{1}{2\pi}\int d\omega \frac{2\pi A(\omega)}{2 \sinh \frac{\beta\omega}{2}}e^{-i\omega t}.
\end{split}
\end{equation}
Therefore, in the frequency domain the Wightman function is
\begin{equation}
G_W(\omega)=\frac{\pi A(\omega)}{\sinh \frac{\beta \omega}{2}}.
\end{equation}

Below we use these various correlators for both rotor $n$ fields and auxiliary $\lambda$ fields. We can associate to each a Euclidean correlator, a retarded correlator, and a Wightman function. They will be distinguished by adding a superscript or subscript to indicate the relevant field.

\section{Four point function}
\subsection{General prescription}

\begin{figure}[H]
\centering
\includegraphics[width=9cm]{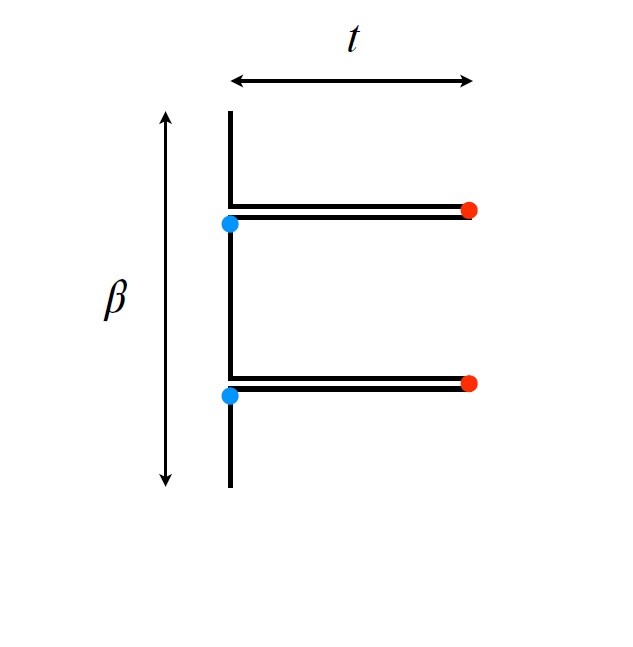}
\caption{Complex time contour defining $\tilde{\mathcal{C}}$. The horizontal direction represents real time while the vertical line is the periodic thermal circle. }
\label{contour}
\end{figure}

This section contains the main analysis of the four-point function of $n$ fields which yields the squared commutator. The squared commutator is
\begin{equation}
\mathcal{C}(t)=-\frac{1}{M^2}\sum_{\mu\nu}\langle [n_i^{\mu}(t),n_j^{\nu}(0)]^2\rangle_{\beta},
\end{equation}
where we have suppressed the position labels $i,j$ on $\mathcal{C}$. To avoid short-time divergences, it's more convenient to consider the regulated version~\cite{brian}:

\begin{equation}
\tilde{\mathcal{C}}(t)=-\frac{1}{M^2}\sum_{\mu\nu} \text{Tr}\left(\rho^{\frac{1}{2}}[n_i^{\mu}(t),n_j^{\nu}(0)]\rho^{\frac{1}{2}}[n_{i\mu}(t),n_{j\nu}(0)]\right),
\end{equation}
This can be interpreted as a combination of contour ordered four point functions living on a complex time contour as in Figure~\ref{contour}.

At large real time $t$, the dominant contribution to $\tilde{\mathcal{C}}(t)$ comes from the propagators stretching between the two horizontal contours. The ones that stretch between the imaginary time line and one of the real time lines cannot affect the large time behavior due to the damping of propagator with respect to the real time separation of the two inserting points. Following this intuition, we will neglect the latter type of dressings to focus on the ladder diagrams shown below.

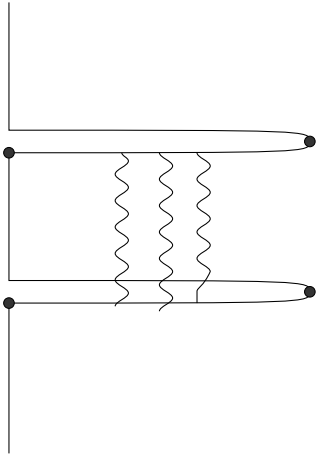
\begin{figure}[H]
\begin{center}
\begin{tikzpicture}
\draw (0,0)--(0,2)..controls (4,2).. (4,2.15)..controls (4,2.3)..(0,2.3)--(0,4)..controls (4,4)..(4,4.15)..controls(4,4.3)..(0,4.3)--(0,6);
\filldraw[fill=black!80] (0,2) circle (2pt)
(4,2.15) circle (2pt) (0,4) circle (2pt) (4,4.15) circle (2pt);
\draw[snake it] (1.5,4)--(1.5,2) (2,4)--(2,2) (2.5,4)--(2.5,2);
\end{tikzpicture}
\end{center}
\caption{Schematic of the ladder diagrams that sum up to give exponential growth of $\tilde{\mathcal{C}}$.}
\end{figure}

For each vertical rung, the real time separation of the two ending points cannot be too large in order for it to remain a finite value. Therefore by integrating their ending points in the real time contour, each vertical rung roughly contributes a factor of $t$. Finally, we also need to sum over all the diagrams with different number of rung insertions. This suggests writing $\tilde{\mathcal{C}}(t)$ in the following form at large $t$,
\begin{equation}\label{eq1}
\tilde{\mathcal{C}}(t) \propto \sum_n \frac{1}{n!}(a t) ^n= e^{at}.
\end{equation}

The factor of $n!$ comes from the permutation of different rungs and $a$ is related to the contribution of the individual rung insertions. In this model, $a$ is a positive number of order $\frac{1}{M}$, which will be clear as we proceed. We will see that a ladder diagram won't damp too fast when two individual rungs are separated by a large real time interval, which justifies the approximation of summing each rung's contribution independently in Eq.~\eqref{eq1}. Similar to the treatment in Ref.~\cite{brian}, we distinguish two types of rungs as shown below.
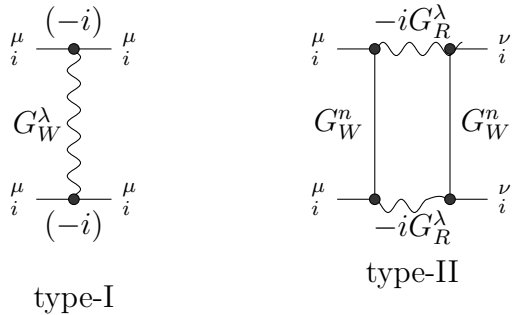
\begin{figure}[H]
\center
\begin{tikzpicture}
\draw[snake it] (-2,0)node[above]{($-i$)}--node[left]{$G_W^{\lambda}$} (-2,-2)node[below]{$(-i)$};
\draw (-2.5,0) node[left]{${}_i^{\mu}$}-- (-2,0) -- (-1.5,0) node[right]{${}_i^{\mu}$} (-2.5,-2) node[left]{${}_i^{\mu}$}--(-2,-2)--(-1.5,-2)node[right]{${}_i^{\mu}$};
\filldraw[fill=black!80] (-2,0) circle (2pt) (-2,-2) circle (2pt);
\draw (-2,-3) node[below]{type-I};
\draw (2,0)--node[left]{$G_W^n$}(2,-2) (3,0)--node[right]{$G_W^n$}(3,-2);
\draw (1.5,0)node[left]{${}_i^{\mu}$}--(2,0) (3,0)--(3.5,0)node[right]{${}_i^{\nu}$} (1.5,-2)node[left]{${}_i^{\mu}$}--(2,-2) (3,-2)--(3.5,-2)node[right]{${}_i^{\nu}$};
\draw[snake it] (2,0)--node[above]{$-iG_R^{\lambda}$}(3,0) (2,-2)--node[below]{$-iG_R^{\lambda}$}(3,-2);
\filldraw[fill=black!80] (2,0) circle (2pt) (3,0) circle (2pt) (2,-2) circle (2pt) (3,-2) circle (2pt);
\draw (2.5,-3) node{type-II} ;
\end{tikzpicture}
\caption{The two types of rungs that enter the ladder summation.}
\label{rung}
\end{figure}

The type-I rung is composed of a single Wightman function of the auxiliary field $\lambda$ stretching between the two real time contours, while the type-II rung is composed of two Wightman functions of rotor fields and two retarded Green's functions of the $\lambda$ field.

Each interaction vertex inserted in the real time contour, which in our case is $(-i)J_{kj}n_k(t)n_j(t)$, has a partner in the other half of the same real time contour (see fig~\ref{vertex} below) but with a minus sign in front. Therefore, adding them together after wick contractions produces a retarded Green's function.

\begin{equation}
\left[\langle n_j(t_2)n_i(t_1)\rangle -\langle n_i(t_1)n_j(t_2)\rangle\right]\theta(t_2-t_1)=(-i)G_R(t_2-t_1)\delta_{ij}
\end{equation}

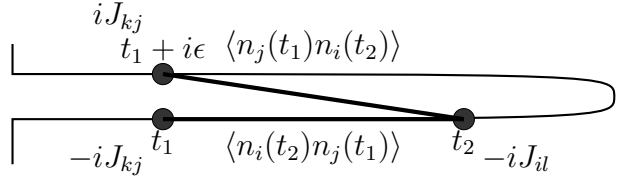
\begin{figure}[H]
\center
\begin{tikzpicture}[scale=2]
\draw[thick] (0,-0.3)--(0,0)..controls (4,0)..(4,0.15)..controls (4,0.3)..(0,0.3)--(0,0.5);
\filldraw[fill=black!80] (1,0) circle (2pt) node[below]{$t_1$} (3,0) circle (2pt) node[below]{$t_2$} (1,0.3) circle (2pt) node[above]{$t_1+i\epsilon$} ;
\draw[ultra thick] (1,0)--node[below]{$\langle n_i(t_2)n_j(t_1)\rangle$}(3,0)  (1,0.3)--(3,0);
\draw (0.95,-0.1)node[below left]{$-iJ_{kj}$} (0.95,0.5)node[above left]{$iJ_{kj}$} (3.05,-0.1)node[below right]{$-iJ_{il}$};
\node[above] at (2,0.3) {$\langle n_j(t_1)n_i(t_2)\rangle$};
\end{tikzpicture}
\caption{An example of interaction vertex insertion. If $n_i(t_2)$ is contracted with $n_j(t_1)$, $n_i$ should at the left of $n_j$ in the correlator. If $n_i(t_2)$ is contracted with $n_j(t_1+i\epsilon)$, they should be reversed due to the contour ordering.  When we add all the possible contractions, these two correlators are combined together to produce a retarded function.}
\label{vertex}
\end{figure}

Therefore, in our following calculations all the propagators that stretch within the same real time contour will be the taken as the retarded Green's function and the ones stretch between two different real time contours are identified as the Wightman function defined in the last section.

\subsection{Leading order}

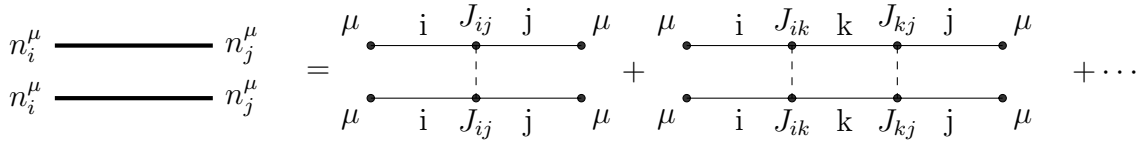
\begin{figure}[H]
    \centering
    \begin{tikzpicture}[scale=0.7]
    \draw[ultra thick] (-6,0)node[left]{$n_i^{\mu}$}--(-3,0)node[right]{$n_j^{\mu}$} (-6,-1)node[left]{$n_i^{\mu}$}--(-3,-1)node[right]{$n_j^{\mu}$};
    \draw[fill=black!80] (0,0) circle (2pt)node[above left]{$\mu$} -- node[above]{i}(2,0) circle (2pt) node[above]{$J_{ij}$}--node[above]{j}(4,0)node[above right]{$\mu$} circle (2pt) (0,-1)node[below left]{$\mu$} circle (2pt) --node[below]{i} (2,-1)  circle (2pt)node[below]{$J_{ij}$}--node[below]{j} (4,-1)node[below right]{$\mu$}circle (2pt);  \draw[fill=black!80] (6,0)node[above left]{$\mu$} circle (2pt) -- node[above]{i}(8,0) circle (2pt) node[above]{$J_{ik}$}--node[above]{k}(10,0) circle (2pt)node[above]{$J_{kj}$}--node[above]{j}(12,0)node[above right]{$\mu$} circle (2pt) (6,-1)node[below left]{$\mu$} circle (2pt) --node[below]{i} (8,-1) circle (2pt)node[below]{$J_{ik}$}--node[below]{k} (10,-1)circle (2pt)node[below]{$J_{kj}$}--node[below]{j} (12,-1)node[below right]{$\mu$}circle (2pt);
    \draw[dashed] (2,0)--(2,-1) (8,0)--(8,-1) (10,0)--(10,-1);
    \draw (-1,-0.5) node{=} (5,-0.5) node{+} (14,-0.5) node{$+\cdots$};
    \end{tikzpicture}
    \caption{Leading order of four point function. The dashed line means the ensemble average of $J_{ij}$. }
    \label{4pt}
\end{figure}

To leading order at large $N$ and $M$, the diagrammatic expansion of the four-point function is shown in Figure~\ref{4pt}. Each disorder average of $J_{ij}$ contains a factor of $\frac{1}{N}$. Each summation over the site indices contains a factor of $N$. So all of these diagrams are of the same order in $\frac{1}{N}$. Since this is just a geometric series, it's easy to obtain that in the leading behavior,
\begin{equation}
\begin{split}
    &\langle n_i^{\mu}(\tau_1)n_i^{\mu}(\tau_2)n_j^{\mu}(\tau_3)n_j^{\mu}(\tau_4)\rangle-\langle n_i^{\mu}(\tau_1)n_i^{\mu}(\tau_2)\rangle \langle n_j^{\mu}(\tau_3)n_j^{\mu}(\tau_4)\rangle \\=&\frac{J^2}{N}T^2\sum_{\omega_n\nu_m}\frac{G(i\omega_n)^2G(i\nu_m)^2}{1-J^2G(i\omega_n)G(i\nu_m)}e^{-i\omega_n\tau_{13}}e^{-i\nu_m \tau_{24}}+(\tau_3\leftrightarrow\tau_4)
\end{split}
\end{equation}

Then we analytically continue it to the real time contour. To do this, just replace $G(i\omega_n)$ by the retarded Green's function $G_R(\omega)$. This is equivalent to summing the diagrams in Figure~\ref{4pt} directly on the real time contour. Either way, the lowest order contribution to $\tilde{\mathcal{C}}(t)$ is
\begin{equation}\label{C0}
-\tilde{\mathcal{C}}_0(t)=\frac{(-iJ)^2}{MN}\int \frac{d\nu}{2\pi}\int\frac{d\omega}{2\pi} \frac{[(-i)G_R(\omega)]^2[(-i)G_R(\nu-\omega)]^2}{1-J^2G_R(\omega)G_R(\nu-\omega)}e^{-i\nu t}.
\end{equation}
After substituting the expression of $G_R$, we find a pole at $\nu=0$,
\begin{equation}\label{pole}
J^2\frac{G_R(\omega)^2G_R(\nu-\omega)^2}{1-J^2G_R(\omega)G_R(\nu-\omega)}\xrightarrow {\nu\rightarrow 0}  \frac{1}{J^2} \frac{\pi}{4}\frac{1}{-i\tilde{\nu}}\frac{JA(\omega)}{\omega}.
\end{equation}
This pole is important for chaos, although it doesn't give the exponential growth by itself. In order to have exponential growth, we must deviate from the strict large $M$ limit and include the rung contributions which are suppressed by $\frac{1}{M}$. Nevertheless, the presence of the pole in the lowest order four-point function allows the rungs to be separated by a large time interval, so that each rung can traverse the whole real time contour independently. Note that the right hand side of Eq.~\eqref{C0} is negative at small $\nu$, but its multiplication with the rungs (Figure~\ref{rung}) is always positive. So the coefficient $a$ in Eq.~\eqref{eq1} is positive.

Now we see that $\frac{1}{M}$ order corrections must be included in the calculation, and therefore we also have to include the self-energy correction to the rotor two point function to that order. This amounts to evaluating the diagram in Figure~\ref{self-energy}, which is computed in the Appendix~\ref{self-energycal}. Write $G_R(\omega)$ as $G^{(0)}_R(\omega)+\frac{1}{M}G^{(1)}_R(\omega)$. Substituting into Eq.~\eqref{pole}, we find that the correction shifts the pole by an amount proportional to $\frac{1}{M}\Re[G^{(0)}_R(\omega)G_R^{(1)}(-\omega)]$, suppressed by a factor of $\frac{1}{M}$.

\begin{figure}
    \centering
    \feynmandiagram [layered layout, horizontal=a to d]{
    a--b--c--d,
    b--[photon, half left]c
    };
    \caption{Diagram corresponding to the rotor self-energy.}
    \label{self-energy}
\end{figure}
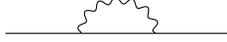

 \subsection{Summing ladder diagrams}

\begin{figure}[H]
    \centering
    \begin{tikzpicture}
    \def \a {4}
    \def \b {7}
    \draw[ultra thick] (0,0)--(3,0) (0,-1)--(3,-1) (\a,0)--(\a+2,0)  (\a,-1)--(\a+2,-1) (\b,0)--(\b+3,0) (\b,-1)--(\b+3,-1);
    \draw[fill=black!20, ultra thick] (1,-1) rectangle (2,0)  (\b+1,0) rectangle (\b+2,-1);
    \draw[snake it] (\b+0.5,0)--(\b+0.5,-1) ;
    \draw (3.5,-0.5)node{=} (\a+2.5,-0.5)node{+};
    \end{tikzpicture}
    \caption{Bethe-Saltpeter equation. The wavy line contains both type-I and type-II rungs.}
    \label{BSeq}
\end{figure}
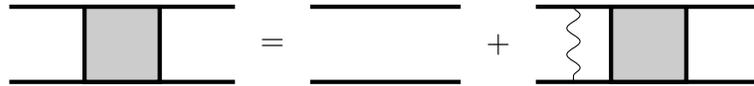

Now we have all the ingredients to calculate the four-point function $\tilde{\mathcal{C}}(t)$ by summing all the ladder diagrams with rungs composed of $\lambda$ propagators. The summation of these diagrams can be done by solving a self-consistent equation. Following the method in Ref.~\cite{brian}, we write down the Bethe-Saltpeter equation for $\tilde{\mathcal{C}}(t)$ as shown in Figure~\ref{BSeq}.

The diagram without any rung insertion is just $\tilde{\mathcal{C}}_0(t)$. Define $f(\nu,\omega)$ by
\begin{equation}
\tilde{\mathcal{C}}(\nu)=\frac{1}{MN}\int \frac{d\omega}{2\pi}f(\nu,\omega).
\end{equation}
Then following Figure~\ref{BSeq}, we can write
\begin{equation}\label{f}
    f(\nu,\omega)=f_0(\nu,\omega)\left[1+\frac{1}{M}\int \frac{d\omega'}{2\pi} G_{\text{rung}}(\omega,\omega')f(\nu,\omega')\right].
\end{equation}

Keeping the corrections up to order $\frac{1}{M}$ in $f_0(\nu,\omega)$ and restricting to small $\nu$ comparable to $\frac{1}{M}$, we approximate $f_0(\nu,\omega)$ by
\begin{equation}\label{f0}
f_0(\nu,\omega)\xrightarrow{\nu\rightarrow 0}
\frac{1}{J^2}\frac{\pi}{4}\frac{1}{(-i\tilde{\nu})\frac{\omega}{JA(\omega)}-\frac{1}{M}\frac{\pi}{2} J^2\Re[G_R^{(0)}(\omega)G^{(1)}_R(-\omega)]}.
\end{equation}
In Eq.~\eqref{f}, $G_{\text{rung}}(\omega,\omega')$ contains the contributions of both type-I and type-II rung. More specifically, it can be written as (see Figure~\ref{rung})
\begin{equation}\label{Grung}
\frac{1}{M}G_{\text{rung}}(\omega,\omega')=\frac{1}{M}\left[\tilde{G}_W^{\lambda}(\omega-\omega')+\int \frac{d\omega''}{2\pi} \tilde{G}_W^n(\omega-\omega'')G_R^{\lambda}(\omega'')\tilde{G}_W^n(\omega''-\omega')G_R^{\lambda}(\nu-\omega'')\right].
\end{equation}

In  Eq.~\eqref{Grung} the wavy line on top of $\Sigma$ and $G$  means that we have extracted the factor $\frac{1}{M}$ from them. Note that the second term of Eq.~\eqref{Grung} contains two $G_R^{\lambda}$'s, but only one $\frac{1}{M}$ is in front. This is because the $O(M)$ indices on the two sides of type-II rung can be different (see fig \ref{rung}) and one of the indices must be summed over when attached to a ladder diagram, which gives a factor of $M$. In summary, in the small $\nu$ limit we rewrite Eq.~\eqref{f} as
\begin{equation}\label{main}
(-i\tilde{\nu})f(\nu,\omega)=\frac{1}{J^2}\frac{\pi}{4}\frac{JA(\omega)}{\omega}\left[1+\frac{1}{M}\int \frac{d\omega'}{2\pi} G_{\text{rung}}(\omega,\omega')f(\nu,\omega')+\frac{J^4}{M}2 \Re[G^{(0)}_R(-\omega)G^{(1)}_R(\omega)]f(\nu,\omega)\right].
\end{equation}

Since $A(\omega)$ is non-zero only in a region $|\tilde{\omega}|\in \mathcal{I}=\left[\sqrt{\lambda'-\frac{1}{2}},\sqrt{\lambda'+\frac{1}{2}}\right] $, we can approximately set $f(\nu,\omega)$ to be zero outside this region. As a result, the integral in terms of $\omega'$ is only over a finite interval, which allows us to discretize the integral and treat it as a matrix multiplication. Note also that by multiplying the both sides by $J^2$ the equation becomes dimensionless. Moreover, we can drop the coupling constant $J$ in the equation and in all the Green's functions appearing in the equation, while replacing all the frequencies and temperature by normalized versions, i.e. $\omega\rightarrow \tilde{\omega}=\frac{\omega}{2\sqrt{gJ}}$. Then the equation is in terms of normalized quantities and all dimensionful couplings are gone (including factors of $2\sqrt{Jg}$). We solve the corresponding eigenvalue problem numerically and restore the physical dimensions of the chaos exponent at the final step, by multiplying it by $2\sqrt{gJ}$. The details are given in Appendix~\ref{numer}.

\section{Discussion and conclusion}

In this section, we discuss the relation between chaos exponent and temperature in some special cases. Of particular interest is the situation near the phase transition between paramagnetic phase and spin glass phase. For simplicity, we only discuss the chaos behavior within the paramagnetic phase and leave the discussion of spin glass phase to the future. We also discuss the pattern of conformal symmetry breaking as compared to the SYK model.

In the future, it would be interesting to study the spin glass phase itself, to understand the interplay between glassiness and quantum information dynamics. It would also be interesting to study the experimental realization of the model, perhaps in a cavity QED or trapped ion setting. The model is experimentally interesting because its relatively analytical tractability makes it a useful benchmark, but it also displays the physics of many-body chaos in the right limit. Another direction building on our work here is to attempt to develop other models with the needed pattern of conformal symmetry breaking to achieve maximal chaos.

\subsection{On the critical line}

\begin{figure}[H]
    \centering
    \includegraphics[scale=0.5]{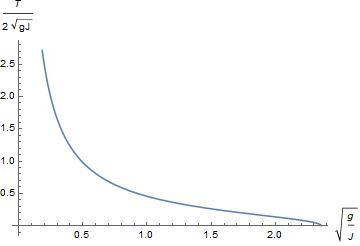}
    \caption{Phase transition line as a function of $\sqrt{g/J}$ and $T/2\sqrt{gJ}$.}
    \label{fig:phase}
\end{figure}

There are three parameters in the model, $T$, $J$ and $g$. By tuning $J$ and $g$ we can reach the phase transition where the parameters obey
\begin{equation}\label{constraint}
2\sqrt{\frac{g}{J}}\int_{0}^1  d\tilde{\omega} JA(\tilde{\omega}) \coth \frac{\tilde{\beta}\tilde{\omega}}{2} = 1
\end{equation}
with
\begin{equation}\label{simple}
    JA(\tilde{\omega})=\frac{2}{\pi}\tilde{\omega}\sqrt{1-\tilde{\omega}^2}
\end{equation}
and $\tilde{\omega} = \frac{\omega}{2\sqrt{gJ}}$. We plot the transition line relating $T$ and $g$, $J$ in Figure~\ref{fig:phase}.

Along the critical line the system is gapless and the spectral function of rotor field takes the simple form Eq.~\eqref{simple}. So it's easier to first discuss the chaos behavior along this critical line. Using the numerical method we can only obtain the chaos exponent for temperatures bigger than the step size of the discretized frequency interval $\mathcal{I}$. Fortunately, for extremely low temperature, we can obtain a more precise relation between the chaos exponent and temperature using some approximations. When $T$ is small, the main equation Eq.~\eqref{main} simplifies. First, the type-I rung contribution is
 \begin{equation}
\begin{split}
    &\int \frac{d\tilde{\omega}'}{2\pi}G_W(\tilde{\omega}-\tilde{\omega}')f(\tilde{\omega}',\tilde{\nu})\\
    =\frac{1}{M}&\int d\tilde{\omega}' \frac{A_{\lambda}(\tilde{\omega}-\tilde{\omega}')}{2\sinh[\frac{1}{2}\tilde{\beta}(\tilde{\omega}-\tilde{\omega}')]}f(\tilde{\omega}',\tilde{\nu})\\
    =\frac{1}{M}&\int \frac{d\bar{\omega'}}{\tilde{\beta}} \frac{A_{\lambda}(\frac{\bar{\omega}-\bar{\omega'}}{\tilde{\beta}})}{2\sinh[\frac{1}{2}(\bar{\omega}-\bar{\omega}')]}f(\tilde{\omega}',\tilde{\nu}), \ \ \ \bar{\omega}=\tilde{\beta}\tilde{\omega}.\\
\end{split}
\end{equation}
At the limit $\tilde{\beta}\rightarrow \infty$, this integral is dominated by $(\bar{\omega}-\bar{\omega'})\sim O(1)$ due to the function $\sinh(\frac{1}{2}(\bar{\omega}-\bar{\omega'}))$ in the denominator. Thus we keep only the leading terms of the expansion of $A_{\lambda}$ in Appendix~\ref{lambdafield},
\begin{equation}\label{lambspecled}
A_{\lambda}\left(\frac{x}{\beta}\right)=\frac{1}{\Pi(0)^2}\frac{1}{\pi^2}\left[\frac{1}{3}\left(\frac{x}{\tilde{\beta}}\right)^3+\frac{4}{3}\pi^2\tilde{T}^2 \left(\frac{x}{\tilde{\beta}}\right)+\cdots\right].
\end{equation}
Note that we have dropped all the dimensionful parameters in the calculation.

To estimate the maximal eigenvalue of this integral kernal, we approximate  $f(\tilde{\omega'})$ by $f(\tilde{\omega})$ and perform the integration, which gives
\begin{equation}
\frac{1}{M}\frac{2\pi^2\tilde{T}^4}{\Pi(0)^2}f(\tilde{\omega},\tilde{\nu}).
\end{equation}

Similarly, the type-II rung contribution can be written as
\begin{equation}\label{typII}
\begin{split}
&M\int \frac{d\tilde{\omega}'}{2\pi}\frac{d\tilde{\omega}''}{2\pi}G_R^{\lambda}(\tilde{\omega}'')G_R^{\lambda}(-\tilde{\omega}'')G_W^n(\tilde{\omega}-\tilde{\omega}'')G_W^n(\tilde{\omega}''-\tilde{\omega}')f(\tilde{\omega}',\tilde{\nu})\\
=&\frac{1}{\tilde{\beta}^2}M\int dx \int dy G_R^{\lambda}\left(\tilde{\omega}+\frac{x}{\tilde{\beta}}\right)G_R^{\lambda}\left(-\tilde{\omega}-\frac{x}{\tilde{\beta}}\right)\frac{A(\frac{x}{\tilde{\beta}})}{2\sinh(\frac{1}{2}x)}\frac{A(\frac{y}{\tilde{\beta}})}{2\sinh(\frac{1}{2}y)}f\left(\tilde{\omega}+\frac{x}{\tilde{\beta}}+\frac{y}{\tilde{\beta}},\tilde{\nu}\right).
\end{split}
\end{equation}

We approximate $G_R(\tilde{\omega}+\frac{x}{\tilde{\beta}})$ by $G_R(\tilde{\omega})$, but still keep the $x$ and $y$ dependence of $f(\tilde{\omega}+\frac{x}{\tilde{\beta}}+\frac{y}{\tilde{\beta}},\tilde{\nu})$ because later we will see that the eigenfunction of interest has to change rapidly near $\tilde{\omega}$. By extracting the temperature  factors, we can see the $T^4$ dependence of type-II rung contribution:
\begin{equation}\label{typeIIscaling}
\tilde{T}^4(\frac{2}{\pi})^2MG_R^{\lambda}(\tilde{\omega})G_R^{\lambda}(-\tilde{\omega})\int{d\tilde{\omega'}}\int_{-\infty}^{\infty} dx \frac{x(\bar{\omega}-\bar{\omega'}-x)}{2\sinh(\frac{1}{2}x)2\sinh(\frac{1}{2}(\bar{\omega}-\bar{\omega'}-x))}f(\tilde{\omega'},\tilde{\nu})
\end{equation}

Finally, we analyze the self-energy contribution. By taking imaginary part of Eq.~\eqref{sigmaR}, we obtain
\begin{equation}\label{Imsigma}
\frac{1}{\pi}\Im\Sigma_R(\omega)=\int dx A_{\lambda}(x)A(\omega-x)(n_B(x)-n_B(x-\omega))
\end{equation}

This function is non-negative for positive frequency. By the results in Appendix~\ref{self-energycal}, we know that the self-energy contribution in Eq.~\eqref{main} is
\begin{equation}
2\Re(G^{(0)}_R(\omega) G^{(1)}_R(-\omega))=-\frac{\Im\Sigma_R(\omega)}{\Im G^{(0)}_R(\omega)}, \ \ \     \text{for $|\omega| \in \mathcal{I}$},
\end{equation}
so it is always non-positive.

At small frequency and low temperature, Eq.~\eqref{Imsigma} has the expansion
\begin{equation}
    \frac{1}{\pi}\Im\Sigma_R(\tilde{\omega})=\frac{1}{\Pi(0)^2}\left(\frac{1}{30\pi^3}\tilde{\omega}^5+\frac{2}{3\pi}\tilde{\omega}^3\tilde{T}^2+\frac{32}{15}\pi \tilde{T}^4\tilde{\omega}+...\right).
\end{equation}
The $\tilde{T}^4$ term dominates at small frequency. Since the positive contributions to chaos exponent are all proportional to $\tilde{T}^4$ at low temperature, we require the eigenfunction to center around $\tilde{\omega}=0$ and decay rapidly within  $O(\frac{1}{\tilde{\beta}})$.

Combining the type-I and type-II rung contributions, we obtain the chaos exponent at very low temperature:
\begin{equation}
\tilde{\lambda}_c=\frac{1.1\times 10^3}{M}\tilde{T}^4
\end{equation}
and
\begin{equation}
\lambda_c=2\sqrt{gJ}\tilde{\lambda}_c=\frac{138}{M}\frac{T^4}{(gJ)^\frac{3}{2}}.
\end{equation}

Here we comment that the $T^4$ dependence of chaos exponent for low temperature still hold for finite value of $M$, while the proportional constant in front of $T^4$ does depend on $M$. As commented in \cite{sachdev}, the higher order of $\frac{1}{M}$ corrections don't modify the scaling dimension of rotor field, so the leading non-analytic part of $G(i\omega_n)$ is always proportional to $|\omega_n|$. Then further check shows that, for $\lambda$ field two point function (Eq.~\eqref{lambdareal}), these corrections only modify the constant $c$, while leave the scaling dimension unchanged. Therefore, $G_{\lambda}(i\omega_n)$ still have it's leading non-analytic part proportional to $|\omega_n|^3$ at critical point. This guarantees the form of Eq.~\eqref{lambspecled} and Eq.~\eqref{typeIIscaling} up to a proportional constant. Then by a similar analysis of this section, one can show that the chaos exponent must be proportional to $T^4$ at low temperature for any value of $M$.

At higher temperature, numerical results show that the chaos exponent always increases with $T$. It saturates at some fixed number when $T$ goes to infinity.

\begin{equation}
\lim_{T\rightarrow\infty}\lambda_c=\frac{5.5\sqrt{gJ}}{M}
\end{equation}

The following table shows the results for finite temperature.
\begin{table}[H]
    \centering
    \begin{tabular}{c|c}
     \textbf{Inverse $\tilde{T}$}    &\textbf{Chaos exponent}  \\
    $\tilde{\beta}$  & $M\tilde{\lambda}_c$ \\
    \hline
    0.01 & 2.75 \\
    0.5 & 2.72 \\
    1 & 2.63\\
    10 & 0.14\\
    30& 0.0014\\
    60& 0.000073\\
    \end{tabular}
    \caption{Numerical data: along the phase transition line. In the table, $\tilde{\beta}=\frac{2\sqrt{gJ}}{T}$, and $\tilde{\lambda_c}=\frac{\lambda_c}{2\sqrt{gJ}}$.}
    \label{numer gapless}
\end{table}

\subsection{At fixed ratio of $\frac{g}{J}$}
Another case we studied is when $\frac{g}{J}$ is fixed and temperature changes. In this case, Eq.~\eqref{constraint} is solved by the following ansatz,
\begin{equation}
    JA(\tilde{\omega})=\frac{2}{\pi}\text{sgn}(\omega)\sqrt{(\tilde{\omega}^2-\lambda'+\frac{1}{2})(\lambda'+\frac{1}{2}-\tilde{\omega}^2)}.
\end{equation}
At low temperature, we obtain the following condition by working with leading order approximation.

\begin{equation}\label{Tcondition}
\delta\lambda \ln \tilde{T}  + \frac{2\pi^2}{3} \tilde{T}^2=0
\end{equation}
where $\delta\lambda=\lambda'-\frac{1}{2}$. The thermal gap as a function of $T$ is determined by it since $\Delta(T)=2\sqrt{gJ\delta\lambda}$. Unlike the previous case, the small frequency and low temperature expansion for spectral functions are more complicated, depending on the relative order of $\omega$ and $T$.

\begin{equation}\label{lambspectral}
\begin{split}
A_{\lambda}(\tilde{\omega})=&\left(\frac{1}{\Pi(0)}\right)^2\frac{1}{\pi^2}\left[\frac{1}{3}\tilde{\omega}^3+2\tilde{\omega}\delta\lambda \ln \tilde{T}+\frac{4\pi^2}{3} \tilde{T}^2\tilde{\omega}\right]+\cdots,  \ \ \ \text{for $\omega\gg T$}\\
=&\left(\frac{1}{\Pi(0)}\right)^2\frac{1}{\pi^2}\left[\frac{1}{3}\tilde{\omega}^3+\frac{4\pi^2}{3} \tilde{T}^2\tilde{\omega}\right]+\cdots,  \ \ \ \ \ \ \ \ \ \ \ \ \ \ \ \ \ \text{for $\omega\ll T$}.
\end{split}
\end{equation}

We can see from this expression that the chaos exponent is still proportional to $T^4$ at low temperature. Then we present some results for general $T$.

\begin{table}[H]
    \centering
    \begin{tabular}{c|c}
     \textbf{Inverse $\tilde{T}$}    &\textbf{Chaos exponent}  \\
    $\tilde{\beta}$  & $M\tilde{\lambda}_c$ \\
    \hline
    0.1 & 121.97 \\
    1  &  4.27\\
    5 & 0.20 \\
    10 & 0.037\\
    30& 0.00058\\
    60& 0.000037\\
    \end{tabular}
    \caption{Relation between chaos exponent and inverse temperature, with $J$ and $g$ fixed. In the table, $\tilde{\beta}=\frac{2\sqrt{gJ}}{T}$, and $\tilde{\lambda_c}=\frac{\lambda_c}{2\sqrt{gJ}}$.}
    \label{numer gapped}
\end{table}

\subsection{Dependence on coupling constant $J$}

We investigated the low temperature limit of the chaos exponent in the previous two sections. From the results, it seems that $\lambda_c$ is negatively related to $J$. However, this is not true since in both two cases we required the ratio $\frac{g}{J}$ to be a constant which may depends on $T$, thus forcing $g$ to increase with $J$. From the Lagrangian we know that $g$ represents the relative importance of quantum effect and it's increase tends to negatively affects the chaos exponent. So in this section we examine numerically the effect of increasing $J$ with $g$ fixed.

The chaos exponent as a function of $T$, $\sqrt{Jg}$ and $\sqrt{\frac{g}{J}}$ is represented by $2\sqrt{gJ}\tilde{\lambda}_c\left(\frac{T}{\sqrt{Jg}},\sqrt{\frac{g}{J}}\right)$. It changes to  $2\sqrt{k}\sqrt{gJ}\tilde{\lambda}_c\left(\frac{1}{\sqrt{k}}\frac{T}{\sqrt{Jg}},\frac{1}{\sqrt{k}}\sqrt{\frac{g}{J}}\right)$ as $J$ changes to $kJ$.  The numerical results show that the new exponent is always smaller than the old one if $k$ is smaller than one, within paramagnetic phase. This matches with the intuition that chaos exponent increases with the magnitude of the random interaction.

\begin{table}[H]
    \centering
    \begin{tabular}{c|c|c}
     \textbf{Inverse $\tilde{T}$}  &\textbf{factor}  &\textbf{Chaos exponent}  \\
    $\tilde{\beta}$ &$k$ & $\sqrt{k}\tilde{\lambda}_c$ \\
    \hline
    &1&2.63\\1 &0.8&2.17\\ & 0.6 &1.68\\
    \hline
     & 1& 0.14\\10 &0.8&0.055\\& 0.6 &0.025\\
     \hline
     & 1& 0.0014\\30 &0.8&0.000018\\&0.6 &$1.3\times10^{-6}$\\
    \end{tabular}
    \caption{Relation between chaos exponent and coupling constant $J$, with $g$ and $T$ fixed. Decreasing $k$ denotes the decreasing of coupling $J$.  }
    \label{numerk}
\end{table}

\subsection{Pattern of conformal symmetry breaking}

The crucial feature of the SYK model that produces maximal chaos is the explicit as well as spontaneous breaking of reparameterization symmetry. According to Ref.~\cite{Maldacena:2016hyu}, this symmetry breaking pattern implies the existence of a pseudo-Goldstone mode with a low energy effective action suppressed by the large coupling constant $\beta J$. It is these pseudo-Goldstone modes that give an enhanced contribution proportional to $\beta J$ in the four point function, which saturates the chaos bound.

In the model that we study, by taking large $\beta\sqrt{Jg}$ limit, one also see invariance of Eq.~\eqref{SDE} under reparameterization transformations of form
\begin{equation}\label{reparametrization}
\begin{split}
    &G(\tau_1,\tau_2)\rightarrow\left[\frac{df(\tau_1)}{d\tau_1}\frac{df(\tau_2)}{d\tau_2}\right]^{\frac{1}{2}}G(f(\tau_1),f(\tau_2)),\\ &\Sigma(\tau_1,\tau_2)\rightarrow \left[\frac{df(\tau_1)}{d\tau_1}\frac{df(\tau_2)}{d\tau_2}\right]^{\frac{1}{2}}\Sigma(f(\tau_1),f(\tau_2))
\end{split}
\end{equation}

However, there exists a function, namely $\delta(\tau_1-\tau_2)$, that is invariant under such transformation. We can see that by taking $\tilde{\omega}_n$ to $0$, the solution Eq.~\eqref{solution} in frequency space indeed reduces to a constant. Therefore, in this model, although the reparametrization symmetry is explicitly broken by the parameter $(\beta\sqrt{gJ})^{-1}$, it is not spontaneously broken to $SL(2,R)$ when  $\beta\sqrt{gJ}$ goes to infinity. So it's not surprising that the chaos exponent is not maximal in this model.  On the other hand, the subleading contribution of order $(\beta\sqrt{gJ})^{-1}$ to the two point function is invariant under $SL(2,R)$ as shown in Eq.~\eqref{conformalrotor}. This is required by the conformal symmetry of the fixed point. Note that the conformal dimension is not the same as appeared in Eq.~\eqref{reparametrization}.

\textit{Acknowledgements:} This work is supported by the Simons Foundation via the It From Qubit collaboration and the National Science Foundation via the Physics Frontier Center at the Joint Quantum Institute. We thank J. Steinberg, S. Xu, and D. Chowdhury for discussions. A related independent work by D. Mao, D. Chowdhury, and T. Senthil will appear at a later date.

\appendix

\section{Propagator for $\lambda$ field}\label{lambdafield}
Since there is no dynamical terms for the auxiliary field $\lambda$ in the Lagrangian, the leading contribution to the $\lambda$ propagator must comes from a loop correction as in Figure~\ref{bubble}. The polarization function is
\begin{equation}\label{pi}
\begin{split}
    M\Pi(i\omega_n)=\sum_{\mu=1}^M\frac{1}{2}&T\sum_{\nu_n}G(i\nu_n)G(-i\nu_n+i\omega_n)\\
    =&M\frac{1}{2}\oint \frac{dz}{2\pi i}\int dx\frac{A(x)}{z-x}\int dy\frac{A(y)}{-z+i\tilde{\omega}_n-y}n_B(z)\\
    =&M\frac{1}{2}\int dx\int dy\frac{A(x)A(y)}{i\tilde{\omega}_n-x-y}[n_B(-y)-n_B(x)]
\end{split}
\end{equation}

$A(x)$ is the spectral function for the rotor field. So the integration is performed on the region where $A(x)A(y)$ is non-zero. In particular, at critical point $A(\tilde{\omega})=\frac{2}{\pi J}\tilde{\omega}\sqrt{1-\tilde{\omega}^2}$, with $\tilde{\omega}\in[-1,1]$. In general, the expression above cannot be simplified further, but if we look at the  strong coupling limit, which amounts to taking $\tilde{\omega}$ to be small, an analytic expression can be obtained. First, we rewrite Eq.~\eqref{pi} as
\begin{equation}
\Pi(i\omega_n)=-\int dt\frac{f(t)}{i\tilde{\omega}_n-t}
\end{equation}
where
\begin{equation}\label{Cnu}
\begin{split}
f(t)=&\frac{1}{2}\int dx A(x)A(t-x)[n_B(x)-n_B(x-t)]\\
=&\frac{2\sqrt{Jg}}{\pi^2 J^2}\left(\frac{1}{3}\tilde{t}^3+\frac{4}{3}\pi^2\tilde{T}^2\tilde{t}+\cdots\right)
\end{split}
\end{equation}

This gives the non-analytic part of $\Pi(i\omega_n)$,
\begin{equation}
\Pi(i\omega_n)=\frac{2\sqrt{Jg}}{\pi J^2}\left(\frac{1}{3}|\tilde{\omega}_n|^3-\frac{4}{3}\pi^2\tilde{T}^2|\tilde{\omega}_n|+\cdots \right)+ \text{analytic part}.
\end{equation}
The non-analytic part of $G_{\lambda}(i\omega_n)$ can also be expanded in $|\tilde{\omega}_n|$ and $\tilde{T}$,
\begin{equation}\label{Glamb}
\begin{split}
G_{\lambda}(i\omega_n)=&-\frac{1}{M\Pi(i\tilde{\omega}_n)}\\
=&\frac{1}{M}\frac{1}{\Pi(0)^2}\frac{2\sqrt{Jg}}{\pi J^2}(\frac{1}{3}|\tilde{\omega}_n|^3-\frac{4}{3}\pi^2\tilde{T}^2|\tilde{\omega}_n|+\cdots)+ \text{analytic part}.
\end{split}
\end{equation}
Finally, by Fourier transformation we obtain the imaginary time propagator of $\lambda$ field at strong coupling limit,
\begin{equation}
T\sum_n G_{\lambda}(i\omega_n)e^{-i\omega_n\tau}=J^2\frac{c}{\sin^4(\frac{\pi\tau}{\beta})}.
\end{equation}

\section{Self-energy correction}\label{self-energycal}
In this section, we compute the self-energy correction (see Figure~\ref{self-energy}),
\begin{equation}\label{def}
\frac{1}{M}\Sigma(i\omega_n)=T\sum_n G(i\omega_n-i\nu_n)G_{\lambda}(i\nu_n)
\end{equation}
where
\begin{equation}\label{Glambda}
    G_{\lambda}(i\nu_n)=-\frac{1}{M\Pi(i\nu_n)}.
\end{equation}
Note that the summation in Eq.~\eqref{def} is not convergent since $G_{\lambda}$(z) defined in Eq.~\eqref{Glambda}) diverges like $\frac{z^2}{a}$ for some constant $a$ as $z$ goes to $\infty$. So we regularize $G_{\lambda}$ by multiplying to it a factor $\frac{1}{1-\epsilon z^2}$ with a small $\epsilon$. Physically, this means we soften the delta function constraint in Eq.~\eqref{delta} by adding a small quadratic term for $\lambda$ in the Lagrangian
\begin{equation}
Z=\int D\lambda Dn e^{\sum_i\frac{1}{2}(\epsilon a M \lambda_i^2-\lambda_in_in_i)+...}
\end{equation}
and finally take $\epsilon$ to zero. When we convert the summation in Eq.~\eqref{def} to contour integral, the effect of this factor is to introduce two poles at $\pm \frac{1}{\sqrt{\epsilon}}$ in the integrand.  Now we first evaluate the integral around branch cuts of function $G_{\lambda}$(z) and $G(z)$. Denoting this part by $\Sigma_1(i\omega_n)$

\begin{equation}\label{sigma}
\begin{split}
    \frac{1}{M}\Sigma_1(i\omega_n)=&\oint\frac{dz}{2\pi i} G(i\omega_n-z)G_{\lambda}(z)n_B(z)\\
    =&-\oint\frac{dz}{2\pi i} \int dx \frac{A(x)}{z-i\omega_n-x}G_{\lambda}(z)n_B(z)+\frac{2i}{2\pi i}\int dx \Im[G_{\lambda}(x+i0^+)]G(x-i\omega_n)n_B(x)\\
    =& \int dx [A(x)G_{\lambda}(x+i\omega_n)n_B(x)+\frac{1}{\pi}\Im[G_{\lambda}(x+i0^+)]G(x-i\omega_n)n_B(x)]
\end{split}
\end{equation}

In the second line, we integrated around the branch cut of $G(i\omega_n+z)$ and $G_{\lambda}(z)$ respectively. Define
\begin{equation}
\frac{1}{M}A_{\lambda}(x)=\frac{1}{\pi}\Im G_{\lambda}(x+i0^+)
\end{equation}

Then from Eq.~\eqref{sigma}, we have that

\begin{equation}\label{branch}
    \frac{1}{M}\Sigma_{1}(i\omega_n)=\int dx A(x)G_{\lambda}(x+i\omega_n)n_B(x)- \frac{1}{M}\int dx \int dy \frac{A_{\lambda}(x)A(y)n_B(x)}{i\omega_n-x-y}
\end{equation}

The integration in $x$ and $y$ are over regions on real axis, where $A(y)$ and $A_{\lambda}(x)$ are non-zero.  Note that we didn't express $G_{\lambda}$ in terms of $A_{\lambda}(x)$ in the first term. In fact $G_{\lambda}$ cannot be represented by it's spectral function in the usual way, since from Eq.~\eqref{Glambda} $G_{\lambda}(z)$ is divergent as $z\rightarrow \infty$. Then we include in Eq.~\eqref{branch} the contribution from the two poles at $z=\pm \frac{1}{\sqrt{\epsilon}}$. Denoting this part by $\Sigma_2$, we have

\begin{equation}\label{poles}
\begin{split}
\frac{1}{M}\Sigma_2(i\omega_n)=&\frac{1}{2\sqrt{\epsilon}}G\left(i\omega_n-\frac{1}{\sqrt{\epsilon}}\right)G_{\lambda}\left(\frac{1}{\sqrt{\epsilon}}\right)n_B\left(\frac{1}{\sqrt{\epsilon}}\right)-\frac{1}{2\sqrt{\epsilon}}G\left(i\omega_n+\frac{1}{\sqrt{\epsilon}}\right)G_{\lambda}\left(-\frac{1}{\sqrt{\epsilon}}\right)n_B\left(-\frac{1}{\sqrt{\epsilon}}\right)\\
\xrightarrow{\epsilon\rightarrow 0}&\frac{1}{2\sqrt{\epsilon}}G\left(\frac{1}{\sqrt{\epsilon}}\right)G_{\lambda}\left(\frac{1}{\sqrt{\epsilon}}\right)\coth\left(\frac{\beta}{2}\frac{1}{\sqrt{\epsilon}}\right)+\frac{i\omega_n}{2\sqrt{\epsilon}}G'\left(\frac{1}{\sqrt{\epsilon}}\right)G_{\lambda}\left(\frac{1}{\sqrt{\epsilon}}\right)
\end{split}
\end{equation}

Since $\frac{1}{\sqrt{\epsilon}}$ is very large, the $\coth$ function can be set to one. So the leading order term is proportional to $\frac{1}{\sqrt{\epsilon}}$ and divergent as $\epsilon\rightarrow 0$. Adding Eq.~\eqref{branch} and Eq.~\eqref{poles} we get
\begin{equation}\label{sigmaR}
\begin{split}
    \frac{1}{M}\Sigma_R(\omega)=&\int dx A(x)G_{\lambda}(x+\omega+i0^+)n_B(x)- \frac{1}{M}\int dx \int dy \frac{A_{\lambda}(x)A(y)n_B(x)}{\omega-x-y+i0^+}\\
&+ \lim_{\epsilon\rightarrow 0} \frac{1}{2\sqrt{\epsilon}}G'\left(\frac{1}{\sqrt{\epsilon}}\right)G_{\lambda}\left(\frac{1}{\sqrt{\epsilon}}\right)\omega+O\left(\frac{1}{M\sqrt{\epsilon}}\right).
\end{split}
\end{equation}

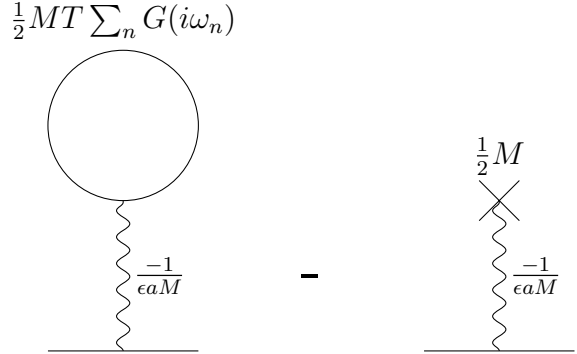
\begin{figure}[H]
    \centering
    \begin{tikzpicture}
    \def \a {5}
    \draw (0,0)--(1,0)--(2,0);
    \draw[snake it] (1,0)--node[right]{$\frac{-1}{\epsilon a M}$}(1,2);
    \draw (1,3) circle[radius=1];
    \draw (1,4) node[above]{$\frac{1}{2}M T\sum_n G(i\omega_n)$};
    \draw (\a,0)--(\a+1,0)--(\a+2,0);
    \draw[snake it] (\a+1,0)--node[right]{$\frac{-1}{\epsilon a M}$}(\a+1,2);
    \draw (\a+1,2) node[cross]{} (\a+1,2.2) node[above]{$\frac{1}{2}M$};
    \draw (\a-1.5,1) node[scale=2]{-};
    \end{tikzpicture}
    \caption{Other diagrams contributing to the self-energy.}
    \label{other self-energy}
\end{figure}

The divergent part is not a problem, since it will be canceled when we add the self-energy diagrams in Figure~\ref{other self-energy}.

After including the self-energy correction, the Schwinger-Dyson equation for the rotor field two point function is modified to
\begin{equation}
G(i\omega_n)^{-1}=\frac{\omega_n^2}{g}+\lambda-J^2G(i\omega_n)-\frac{1}{M}\tilde{\Sigma}(i\omega_n)
\end{equation}

Solving this equation we obtain

\begin{equation}
\begin{split}
G(i\omega_n)=&G^{(0)}(i\omega_n)+\frac{1}{M}G^{(1)}(i\omega_n)\\
=&G^{(0)}(i\omega_n)+\frac{1}{M}\frac{c-\tilde{\Sigma}(i\omega_n)}{J^2-G^{(0)}(i\omega_n)^{-2}}
\end{split}
\end{equation}
where $\frac{c}{M}=\lambda-\lambda^{(0)}$ is a constant. $\tilde{\Sigma}(i\omega_n)$ is the finite part of $\Sigma(i\omega_n)$ after subtracting the $O(\frac{1}{\sqrt{\epsilon}})$ term.  The pole in Eq.~\eqref{pole} is modified to
\begin{equation}
J^2\frac{G_R(\omega)^2G_R(\nu-\omega)^2}{1-J^2G_R(\omega)G_R(\nu-\omega)}\xrightarrow {\nu\rightarrow 0}  \frac{1}{J^2} \frac{\pi}{4}\frac{JA(\omega)}{\omega}\frac{1}{-i\tilde{\nu}-\frac{\pi}{2}\frac{1}{M}\frac{JA(\omega)}{\omega}J^2 \Re(G_R^{(0)}(-\omega)G_R^{(1)}(\omega))}
\end{equation}
When $\omega$ lies in the interval such that $A(\omega)\neq 0$, we have
\begin{equation}
\begin{split}
\Re(G_R^{(0)}(-\omega)G_R^{(1)}(\omega))=\frac{1}{J^4}& \Re\left(\frac{c-\tilde{\Sigma}_R(\omega)}{G_R(\omega)-G_R(-\omega)}\right)\\
=&\frac{1}{J^4}\frac{-\Im\tilde{\Sigma}_R(\omega)}{2\Im G_R(\omega)}
\end{split}
\end{equation}

This function is always non-positive since both the denominator and numerator are non-negative functions. The constant $c$ need not to be determined for our purposes, but it can be fixed by demanding
\begin{equation}
T\sum_n G(i\omega_n)=1.
\end{equation}

\section{Details of numerical calculation}\label{numer}
In Eq.~\eqref{main}, we redefine $f'(\omega,\nu)=\sqrt{\frac{A(\omega)}{\omega}}f(\omega,\nu)$. Then it is converted to an eigenvalue problem with a symmetric kernel,
\begin{equation}
K(\omega,\omega'):=\frac{\pi}{4}\frac{1}{M}\left[\sqrt{\frac{JA(\omega)}{\omega}}\sqrt{\frac{JA(\omega')}{\omega'}}G_{\text{rung}}(\omega,\omega')+2\frac{JA(\omega)}{\omega}\Re(G^{(0)}_R(-\omega)G^{(1)}_R(\omega))(2\pi)\delta(\omega-\omega')\right]
\end{equation}
and
\begin{equation}\label{kernal}
    \partial_t f'(\omega,t)=\int\frac{d\omega'}{2\pi}K(\omega,\omega')f'(\omega',t).
\end{equation}

In the practical calculations, we only deal with the normalized frequency $\tilde{\omega}:=\frac{\omega}{2\sqrt{gJ}}$. Since $A(\tilde{\omega})$ is non-zero only when $|\tilde{\omega}|$ lies in finite interval $\mathcal{I}=[\sqrt{\lambda'-\frac{1}{2}},\sqrt{\lambda'+\frac{1}{2}}]$, approximately we can say that $f(\omega,\nu)$ is only non-zero on $\mathcal{I}$. Then we discretize this interval and diagonalize the kernel numerically by treating Eq.~\eqref{kernal} as a matrix equation,

\begin{equation}
    \partial_t \hat{f}(t)=\hat{K}\hat{f}(t)
\end{equation}
where $\hat{f}(t)$ is a vector whose components are $f(\tilde{\omega}=\frac{i}{m},t)$. $\hat{K}$ is a matrix with matrix elements $\hat{K}_{ij}=\frac{1}{m}\frac{1}{2\pi}K(\frac{i}{m},\frac{j}{m})$. $\frac{1}{m}$ is the step size. In our calculation, we take it to be $\frac{1}{100}$. At low temperature, we refine the gridding in the region where $n_B(x)$ varies fast, in order to improve the precision. Then we found the maximal eigenvalue $\lambda_m$ of $\hat{K}$ by diagonalizing it using Mathematica. If $\lambda_m$ is positive, the corresponding growing mode will dominate for large $t$. So we identify the chaos exponent $\tilde{\lambda}_c$ as $\lambda_m$.

\bibliographystyle{unsrt}
\bibliography{biblio}

\begin{thebibliography}{10}

\bibitem{Shenker2014}
Stephen~H. Shenker and Douglas Stanford.
\newblock {Black holes and the butterfly effect}.
\newblock {\em JHEP}, 2014(3):67, mar 2014.

\bibitem{Kitaev_talk}
Alexei Kitaev.
\newblock {A simple model of quantum holography, Part 1 and Part 2 (talks at
  KITP, Santa Barbara)}.

\bibitem{Hayden2007}
Patrick Hayden and John Preskill.
\newblock {Black holes as mirrors: quantum information in random subsystems}.
\newblock {\em JHEP}, 2007(9):120, sep 2007.

\bibitem{Sekino2008}
Yasuhiro Sekino and Leonard Susskind.
\newblock {Fast scramblers}.
\newblock {\em JHEP}, 2008(10):065, oct 2008.

\bibitem{Hosur2016}
Pavan Hosur, Xiao-Liang Qi, Daniel~A. Roberts, and Beni Yoshida.
\newblock {Chaos in quantum channels}.
\newblock {\em JHEP}, 2016(2):4, feb 2016.

\bibitem{larkin1969quasiclassical}
AI~Larkin and Yu~N Ovchinnikov.
\newblock Quasiclassical method in the theory of superconductivity.
\newblock {\em Sov Phys JETP}, 28(6):1200--1205, 1969.

\bibitem{Almheiri:2013hfa}
Ahmed Almheiri, Donald Marolf, Joseph Polchinski, Douglas Stanford, and James
  Sully.
\newblock {An Apologia for Firewalls}.
\newblock {\em JHEP}, 09:018, 2013.

\bibitem{Nahum2017a}
Adam Nahum, Sagar Vijay, and Jeongwan Haah.
\newblock {Operator Spreading in Random Unitary Circuits}.
\newblock {\em Phys. Rev. X}, 8(2):021014, apr 2018.

\bibitem{VonKeyserlingk2017}
Curt von Keyserlingk, Tibor Rakovszky, Frank Pollmann, and Shivaji Sondhi.
\newblock {Operator hydrodynamics, OTOCs, and entanglement growth in systems
  without conservation laws}.
\newblock {\em Phys. Rev. X}, 8:021013, 2018.

\bibitem{Xu2018}
Shenglong Xu and Brian Swingle.
\newblock {Accessing scrambling using matrix product operators}.

\bibitem{Sachdev1993}
Subir Sachdev and Jinwu Ye.
\newblock {Gapless spin-fluid ground state in a random quantum Heisenberg
  magnet}.
\newblock {\em Phys. Rev. Lett.}, 70(21):3339--3342, may 1993.

\bibitem{Gu2017}
Yingfei Gu, Xiao-Liang Qi, and Douglas Stanford.
\newblock {Local criticality, diffusion and chaos in generalized
  Sachdev-Ye-Kitaev models}.
\newblock {\em JHEP}, 2017(5):125, may 2017.

\bibitem{Luitz2017}
David~J. Luitz and Yevgeny {Bar Lev}.
\newblock {Information propagation in isolated quantum systems}.
\newblock {\em Phys. Rev. B}, 96(2):020406, jul 2017.

\bibitem{Bohrdt2017a}
A~Bohrdt, C~B Mendl, M~Endres, and M~Knap.
\newblock {Scrambling and thermalization in a diffusive quantum many-body
  system}.
\newblock {\em New J. Phys.}, 19(6):063001, dec 2016.

\bibitem{Heyl2018}
Markus Heyl, Frank Pollmann, and Bal{\'{a}}zs D{\'{o}}ra.
\newblock {Detecting equilibrium and dynamical quantum phase transitions via
  out-of-time-ordered correlators}.
\newblock {\em Phys. Rev. Lett.}, 121:016801, 2018.

\bibitem{Lin2018}
Cheng-Ju Lin and Olexei~I. Motrunich.
\newblock {Out-of-time-ordered correlators in a quantum Ising chain}.
\newblock {\em Phys. Rev. B}, 97(14):144304, apr 2018.

\bibitem{Nahum2017}
Adam Nahum, Jonathan Ruhman, Sagar Vijay, and Jeongwan Haah.
\newblock {Quantum Entanglement Growth under Random Unitary Dynamics}.
\newblock {\em Phys. Rev. X}, 7(3):031016, jul 2017.

\bibitem{Rakovszky2017}
Tibor Rakovszky, Frank Pollmann, and C.~W. von Keyserlingk.
\newblock {Diffusive hydrodynamics of out-of-time-ordered correlators with
  charge conservation}.
\newblock 2017.

\bibitem{Khemani2017}
Vedika Khemani, Ashvin Vishwanath, and David~A Huse.
\newblock {Operator spreading and the emergence of dissipation in unitary
  dynamics with conservation laws}.

\bibitem{Khemani2018}
Vedika Khemani, David~A Huse, and Adam Nahum.
\newblock {Velocity-dependent Lyapunov exponents in many-body quantum,
  semi-classical, and classical chaos}.

\bibitem{Lashkari2012}
Nima Lashkari, Douglas Stanford, Matthew Hastings, Tobias Osborne, and Patrick
  Hayden.
\newblock {Towards the fast scrambling conjecture}.
\newblock {\em JHEP}, 2013(4):22, apr 2013.

\bibitem{Shenker2014a}
Stephen~H. Shenker and Douglas Stanford.
\newblock {Stringy effects in scrambling}.
\newblock {\em JHEP}, 2015(5):132, may 2015.

\bibitem{Aleiner2016}
Igor~L Aleiner, Lara Faoro, and Lev~B Ioffe.
\newblock {Microscopic model of quantum butterfly effect: Out-of-time-order
  correlators and traveling combustion waves}.
\newblock {\em Ann. Phys. (N. Y).}, 375:378--406, 2016.

\bibitem{Swingle:2016jdj}
Brian Swingle and Debanjan Chowdhury.
\newblock {Slow scrambling in disordered quantum systems}.
\newblock {\em Phys. Rev. B}, 95(6):060201, 2017.

\bibitem{Grozdanov2018}
Sa{\v{s}}o Grozdanov, Koenraad Schalm, and Vincenzo Scopelliti.
\newblock {Kinetic theory for classical and quantum many-body chaos}.

\bibitem{Patel2017}
Aavishkar~A Patel, Debanjan Chowdhury, Subir Sachdev, and Brian Swingle.
\newblock {Quantum Butterfly Effect in Weakly Interacting Diffusive Metals}.
\newblock {\em Phys. Rev. X}, 7(3):031047, sep 2017.

\bibitem{Swingle2018Resilience}
Brian Swingle and Nicole Yunger~Halpern.
\newblock Resilience of scrambling measurements.
\newblock {\em Phys. Rev. A}, 97:062113, Jun 2018.

\bibitem{Dressel2018SCweak}
Justin Dressel, Jos\'e~Ra\'ul Gonz\'alez~Alonso, Mordecai Waegell, and Nicole
  Yunger~Halpern.
\newblock Strengthening weak measurements of qubit out-of-time-order
  correlators.
\newblock {\em Phys. Rev. A}, 98:012132, Jul 2018.

\bibitem{Menezes2018Sonic}
G.~{Menezes} and J.~{Marino}.
\newblock {Slow scrambling in sonic black holes}.
\newblock {\em EPL (Europhysics Letters)}, 121:60002, March 2018.

\bibitem{Scaffidi2017Semiclassical}
T.~{Scaffidi} and E.~{Altman}.
\newblock {Semiclassical Theory of Many-Body Quantum Chaos and its Bound}.
\newblock {\em ArXiv e-prints}, November 2017.

\bibitem{swingle2016}
Brian Swingle, Gregory Bentsen, Monika Schleier-Smith, and Patrick Hayden.
\newblock {Measuring the scrambling of quantum information}.
\newblock {\em Phys. Rev. A}, 94(4):040302, oct 2016.

\bibitem{Zhu2016}
Guanyu Zhu, Mohammad Hafezi, and Tarun Grover.
\newblock {Measurement of many-body chaos using a quantum clock}.
\newblock {\em Phys. Rev. A}, 94(6):062329, dec 2016.

\bibitem{Yao2016a}
Norman~Y. Yao, Fabian Grusdt, Brian Swingle, Mikhail~D. Lukin, Dan~M.
  Stamper-Kurn, Joel~E. Moore, and Eugene~A. Demler.
\newblock {Interferometric Approach to Probing Fast Scrambling}.
\newblock 2016.

\bibitem{Halpern2016}
Nicole {Yunger Halpern}.
\newblock {Jarzynski-like equality for the out-of-time-ordered correlator}.
\newblock {\em Phys. Rev. A}, 95(1):012120, jan 2017.

\bibitem{Halpern2017}
Nicole Yunger~Halpern, Brian Swingle, and Justin Dressel.
\newblock {Quasiprobability behind the out-of-time-ordered correlator}.
\newblock {\em Phys. Rev.}, A97(4):042105, 2018.

\bibitem{Campisi2017}
Michele Campisi and John Goold.
\newblock {Thermodynamics of quantum information scrambling}.
\newblock {\em Phys. Rev. E}, 95(6):062127, jun 2017.

\bibitem{Yoshida2017}
Beni Yoshida and Alexei Kitaev.
\newblock {Efficient decoding for the Hayden-Preskill protocol}.

\bibitem{Garttner2016}
Martin G\"arttner, Justin~G. Bohnet, Arghavan Safavi-Naini, Michael~L. Wall,
  John~J. Bollinger, and Ana~Maria Rey.
\newblock {Measuring out-of-time-order correlations and multiple quantum
  spectra in a trapped-ion quantum magnet}.
\newblock {\em Nat. Phys.}, 13(8):781--786, aug 2017.

\bibitem{Wei2016}
Ken~Xuan Wei, Chandrasekhar Ramanathan, and Paola Cappellaro.
\newblock Exploring localization in nuclear spin chains.
\newblock {\em Phys. Rev. Lett.}, 120:070501, Feb 2018.

\bibitem{Li2017a}
Jun Li, Ruihua Fan, Hengyan Wang, Bingtian Ye, Bei Zeng, Hui Zhai, Xinhua Peng,
  and Jiangfeng Du.
\newblock {Measuring out-of-time-order correlators on a nuclear magnetic
  resonance quantum simulator}.
\newblock {\em Phys. Rev. X}, 7(3):031011, jul 2017.

\bibitem{Meier2017}
Eric~J. {Meier}, Jackson {Ang'ong'a}, Fangzhao~Alex {An}, and Bryce {Gadway}.
\newblock {Exploring quantum signatures of chaos on a Floquet synthetic
  lattice}.
\newblock {\em arXiv e-prints}, page arXiv:1705.06714, May 2017.

\bibitem{Landsman2018}
Kevin~A. {Landsman}, Caroline {Figgatt}, Thomas {Schuster}, Norbert~M. {Linke},
  Beni {Yoshida}, Norman~Y. {Yao}, and Christopher {Monroe}.
\newblock {Verified Quantum Information Scrambling}.
\newblock {\em arXiv e-prints}, page arXiv:1806.02807, June 2018.

\bibitem{Wei2018}
Ken~Xuan {Wei}, Pai {Peng}, Oles {Shtanko}, Iman {Marvian}, Seth {Lloyd},
  Chandrasekhar {Ramanathan}, and Paola {Cappellaro}.
\newblock {Emergent prethermalization signatures in out-of-time ordered
  correlations}.
\newblock {\em arXiv e-prints}, page arXiv:1812.04776, December 2018.

\bibitem{Sachdev:2015efa}
Subir Sachdev.
\newblock {Bekenstein-Hawking Entropy and Strange Metals}.
\newblock {\em Phys. Rev. X}, 5(4):041025, 2015.

\bibitem{Maldacena:2016hyu}
Juan Maldacena and Douglas Stanford.
\newblock {Remarks on the Sachdev-Ye-Kitaev model}.
\newblock {\em Phys. Rev. D}, 94(10):106002, 2016.

\bibitem{sachdev}
N.~Read Jinwu~Ye, Subir~Sachdev.
\newblock A solvable spin-glass of quantum rotors.
\newblock {\em Phys. Rev. lett}, 70:4011, 1993.
\newblock \url{https://arxiv.org/abs/cond-mat/9212027}.

\bibitem{2017PTEP.2017h3I01D}
I.~{Danshita}, M.~{Hanada}, and M.~{Tezuka}.
\newblock {Creating and probing the Sachdev-Ye-Kitaev model with ultracold
  gases: Towards experimental studies of quantum gravity}.
\newblock {\em Progress of Theoretical and Experimental Physics},
  2017(8):083I01, August 2017.

\bibitem{2018PhRvL.121c6403C}
A.~{Chen}, R.~{Ilan}, F.~{de Juan}, D.~I. {Pikulin}, and M.~{Franz}.
\newblock {Quantum Holography in a Graphene Flake with an Irregular Boundary}.
\newblock {\em Physical Review Letters}, 121(3):036403, July 2018.

\bibitem{2018arXiv180901671G}
H.~{Gharibyan}, M.~{Hanada}, B.~{Swingle}, and M.~{Tezuka}.
\newblock {Quantum Lyapunov Spectrum}.
\newblock {\em ArXiv e-prints}, September 2018.

\bibitem{boundonchaos}
Douglas~Stanford Juan~Maldacena, Stephen H.~Shenker.
\newblock A bound on chaos.
\newblock {\em JHEP08(2016)106}.
\newblock \url{https://arxiv.org/abs/1503.01409}.

\bibitem{brian}
Brian~Swingle Debanjan~Chowdhury.
\newblock Onset of many-body chaos in the o(n) model.
\newblock {\em Phys. Rev. D}, 96, 065005, 2017.
\newblock \url{https://arxiv.org/abs/1703.02545}.

\end{thebibliography}
\end{document}